\documentclass[
 reprint,
superscriptaddress,
 amsmath,amssymb,
 aps,
prb,
longbibliography
]{revtex4-2}

\usepackage[utf8]{inputenc}
\usepackage[english]{babel}
\usepackage{dcolumn}
\usepackage{graphicx}
\usepackage{amsmath}
\usepackage{amsfonts}
\usepackage{amssymb}
\usepackage{mathtools}
\usepackage{bm}
\usepackage{bbm}
\usepackage[colorlinks=true,allcolors=blue]{hyperref}
\usepackage{siunitx}
\usepackage{amsthm}
\usepackage{cleveref}
\usepackage{color}

\newcommand{\abs}[1]{\left|#1\right|}
\newcommand{\ucav}[1]{\left| \overline{a}_{#1} \right|^2}

\begin{document}

\preprint{APS/123-QED}

\title{Electromagnetically induced transparency from first-order dynamical systems}

\author{Marco Clementi}
\email{marco.clementi01@universitadipavia.it}
\affiliation{Dipartimento di Fisica, Universit\`a degli Studi di Pavia, Via Agostino Bassi 6, I-27100 Pavia, Italy}
\author{Matteo Galli}%
\affiliation{Dipartimento di Fisica, Universit\`a degli Studi di Pavia, Via Agostino Bassi 6, I-27100 Pavia, Italy}
\author{Liam O'Faolain}%
\affiliation{Centre for Advanced Photonics and Process Analysis, Munster Technological University, Rossa Ave Bishopstown, Cork T12 P928, Ireland}
\affiliation{Photonics, Tyndall National Institute, Lee Maltings Complex Dyke Parade, Cork T12 R5CP, Ireland}
\author{Dario Gerace}
 \email{dario.gerace@unipv.it}
\affiliation{Dipartimento di Fisica, Universit\`a degli Studi di Pavia, Via Agostino Bassi 6, I-27100 Pavia, Italy}

\date{\today}

\begin{abstract}
We show how a strongly driven single-mode oscillator coupled to a first-order dynamical system gives rise to induced absorption or gain of a weak probe beam, and associated fast or slow light depending on the detuning conditions. We derive the analytic solutions to the dynamic equations of motion, showing that the electromagnetically induced transparency (EIT) like response is a general phenomenology, potentially occurring in any nonlinear oscillator coupled to first-order dynamical systems. The resulting group delay (or advance) of the probe is fundamentally determined by the system damping rate. To illustrate the practical impact of this general theoretical framework, we quantitatively assess the observable consequences of either thermo-optic or free-carrier dispersion effects in conventional semiconductor microcavities in control/probe experiments, highlighting the generality of this physical mechanism and its potential for the realization of EIT-like phenomena in integrated and cost-effective photonic devices.
\end{abstract}


\maketitle

\section{Introduction}
Electromagnetically induced transparency (EIT) \cite{Fleischhauer2005} and its analogues \cite{Souza2015} gathered high interest over the last few decades. 
In its broadest definition, the phenomenon consists in the appearance of a narrow transmission window in an otherwise absorptive medium, such as an atomic cloud or a solid state medium. 
This is typically associated with a steep normal dispersion, owing to the analytic relation between the real and imaginary parts of the dielectric response function, which gives rise to large group delays (slow light) for light pulses within the induced transparency spectral window \cite{Boyd2009}, with applications ranging from optical delay lines to photonic quantum memories \cite{Lukin2001}.

After the first experimental evidence shown on atomic clouds \cite{Boller1991}, and the subsequent demonstration of slowing light down to \SI{17}{\meter/\second} in an ultracold gas \cite{Hau1999}, several EIT-analogues have been reported in solid-state systems, such as  semiconductor quantum dots \cite{Faraon2008a,Brunner2009}, coupled resonators \cite{Xu2006}, metamaterials \cite{Papasimakis2008}, cavity optomechanics \cite{Weis2010, Safavi-Naeini2011}, and acousto-optic resonators \cite{Dong2015,Kim2015}, with group delays or advances as high as few microseconds.
The possibility to achieve optical delay was specifically investigated in integrated photonics, due to the interest for the implementation of small-footprint delay lines, with applications ranging from optical memories \cite{Alexoudi2020} to quantum computing \cite{Pichler2017}.

All the EIT-analogues quoted above share a common feature: the electromagnetic field coherently interacts with a physical system displaying a second-order dynamic response, characteristic of harmonic oscillators. This may be either a $\Lambda$-type three level system in the case of conventional EIT \cite{Fleischhauer2005}, or a mechanical harmonic oscillator in the case of optomechanically induced transparency (OMIT) \cite{Agarwal2010,Weis2010,Safavi-Naeini2011}, or again an acoustic mode of the structure in the case of Brillouin scattering induced transparency (BSIT) \cite{Dong2015,Kim2015}.
Intuitively, all these examples have in common the coherent exchange of energy between the field and the harmonic degree of freedom, which results in the emergence of the induced transparency window, and the associated slow or fast light effects \cite{Safavi-Naeini2011}.

{In contrast, little effort has been devoted so far to  investigate the interaction between the electromagnetic field and a first-order (i.e., dissipative) systems. In this work,} we present a theoretical model of a  generic single-mode oscillator driven by an intense control and a weak probe fields, and nonlinearly coupled to a first-order dynamical system. We show how this very general physical configuration gives rise to an EIT-analogue response.
In light of the several possible physical realizations of this model, our approach represents a novel framework to investigate EIT and its peculiar consequences on light dispersion in various and unconventional experimental settings.
In particular, we first show that any dynamical quantity characterized by a decay rate $\gamma$ and mediating an effective Kerr-type nonlinear interaction on the localized electromagnetic field gives rise to a spectral hole or anti-hole. These can be experimentally evidenced by a weak probe superimposed to an intense control beam, associated group advance or delay, respectively.

We then show how this phenomenology stems from two paradigmatic examples of effective nonlinearities in semiconductor resonators: the first-order dynamical response associated to thermo-optic effect, and the one due to free-carrier dispersion. 
As a consequence, any resonator realized in material platforms subject to such types of nonlinear behavior can be potentially engineered to display EIT-like response and to exploit their observable consequences, i.e., large group delays or advances. 
We note that, while the first experimental evidences for thermo-optically induced transparency have already been reported \cite{Clementi2020, Ma_SciAdv_2020}, the free-carrier induced transparency phenomenon predicted here is a novel effect.
While this work aims at providing a theoretical framework for the formulation and interpretation of a novel phenomenon, our purpose is also to introduce the key analytic tools to design the practical implementation of large group delays and advances in actual photonic devices based on conventional  material platforms, thus fostering new experiments allowing to leverage the stringent requirements of typical EIT-analogue physics.

The manuscript is organized as follows.
In Sec.~\ref{sec:general_model} we show how EIT can emerge from the interaction between a confined optical mode and a first-order system, within a general theoretical framework.
First, we present the relevant equations of motion, their steady state solution (Sec.~\ref{sec:bistability}), and a linearized solution assuming a control-probe excitation (Sec.~\ref{sec:dynamic_solution}).
In Sec.~\ref{sec:applications}, we predict how this phenomenon can be observed in actual state-of-art semiconductor microcavities, mediated either by thermo-optic effect (Sec.~\ref{sec:toit}) or by free-carrier dispersion effect (Sec.~\ref{sec:fcit}), and we finally provide a comparison of the relevant figures of merit (Sec.~\ref{sec:disc}).

\section{General model}
\label{sec:general_model}

\begin{figure}
    \centering
    \includegraphics[width=0.4\textwidth]{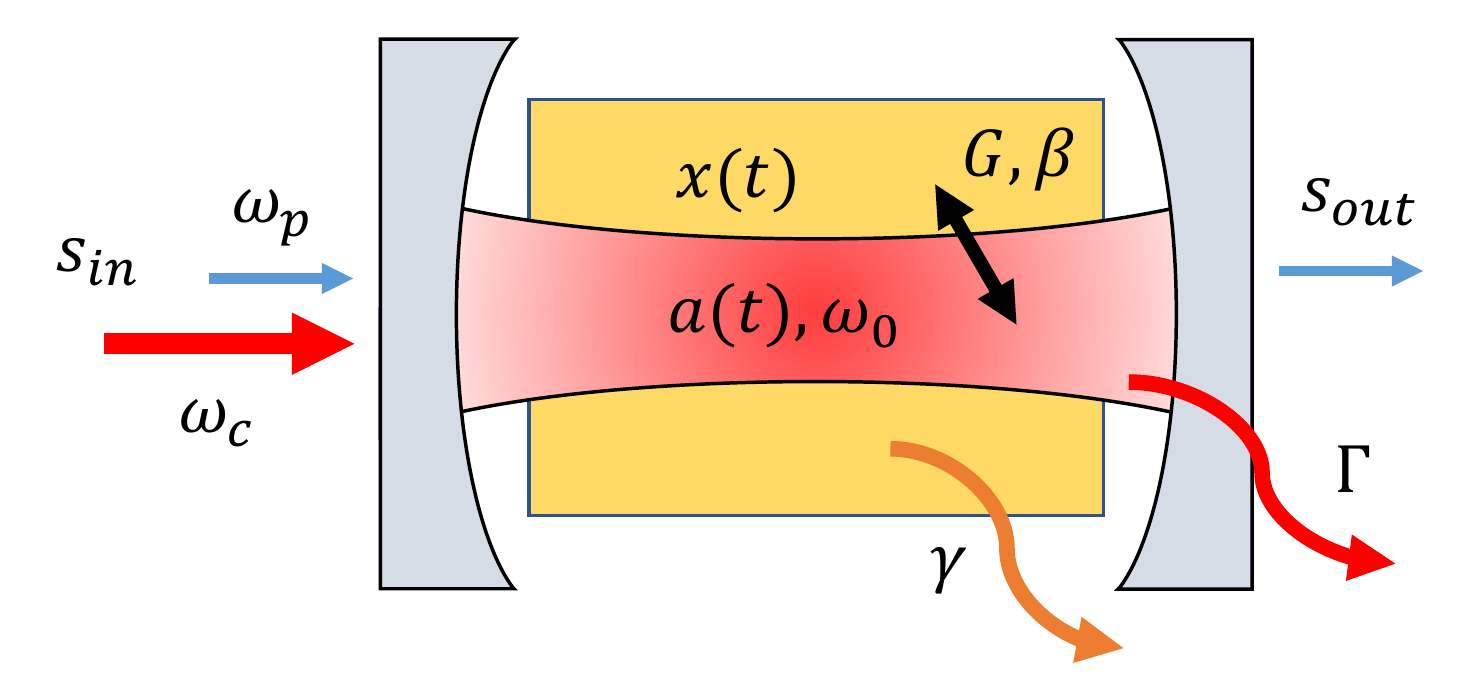}
    \caption{
    Scheme of the physical system under analysis.  
    An optical resonator, characterized by an intracavity field amplitude $a(t)$, resonance frequency $\omega_0$, and intrinsic decay rate $\Gamma$, is coherently driven by a control (red arrow) and a probe (blue arrow) input fields. 
    The confined mode is coupled to a first-order system (yellow), uniquely characterized by a dynamical variable $x(t)$ and an incoherent dissipation rate $\gamma$ through the coupling parameters $G$ and $\beta$.  }
    \label{fig:fig1}
\end{figure}

Consider a single-mode oscillator, which for simplicity might be thought as an optical cavity (e.g., sketched in Fig.~\ref{fig:fig1}), characterized by a time-dependent classical field amplitude ${a(t)}$, a resonance frequency $\overline{\omega}_0$, a decay rate $\Gamma$, and driven by a forcing amplitude $s_{in}$. 
Next, let us consider a second physical system, described by a classical degree of freedom $x(t)$, and obeying the following first-order rate equation:
\begin{equation}
\frac{d}{dt} x(t)=\beta\abs{a(t)}^2 - \gamma x(t)
\label{eq:dynamics_x}
\end{equation}
where the forcing term is proportional to the intracavity energy, $\abs{a(t)}^2$ through the absorptive parameter $\beta$, while the dissipation rate $\gamma$ represents the intrinsic system damping (decay).

We will henceforth assume that the resonance frequency parametrically depends on $ x(t)$, such that $\overline{\omega}_0=\omega_0+G x(t)$, in which $G$ is a coupling term, and $\omega_0$ is the resonance frequency in the absence of any nonlinear shift.
We assume the field amplitude to obey the following dynamical equation:
\begin{equation}
\frac{d}{dt}a(t) =
\left(i\omega_0 - \frac{\Gamma}{2} \right)a(t) + 
i G  x(t) a(t) +
\sqrt{\eta \Gamma} s_{in}(t)
\label{eq:dynamics_field}
\end{equation}
\\
in which $\eta$ represents the in-coupling efficiency to the single-mode oscillator, such that the in-coupling rate is given as $\Gamma_{in}=\eta\Gamma$.
The interaction term in Eq.~\eqref{eq:dynamics_field} can be easily interpreted as a Kerr-type (i.e. intensity-dependent) shift of the cavity resonance frequency, typically observed in nonlinear optical resonators.
However, in contrast with the instantaneous Kerr response, the timescale of such interaction is here governed by the characteristic damping rate, $\gamma$, typical of effective nonlinearities.
In analogy with the usual phenomenology of Kerr response, we will assume $G<0$ for the moment, with straightforward generalization.

We will now study the dynamics of the system response first in the presence of a single driving field with a constant amplitude, and then in a control and probe excitation configuration.

\subsection{Steady state solution: optical bistability}
\label{sec:bistability}
Consider a monochromatic driving field (henceforth referred to as ``control field") in the form:
\begin{equation}
    s_{in}(t)=\overline{s}_{in}e^{i\omega_c t}
    \label{eq:control}
\end{equation}
\\
The steady state solutions to Eqs.~\eqref{eq:dynamics_x} and \eqref{eq:dynamics_field} are:
\begin{equation}
\overline{a} = 
\frac{\sqrt{\eta \Gamma}}
{i\overline{\Delta} + \Gamma/2}
\overline{s}_{in} \qquad
\overline{ x} = \frac{\beta}{\gamma} \ucav{}
\label{eq:steady_state}
\end{equation}
\\
where $\overline{\Delta}=\omega_c-\overline{\omega}_0=\omega_c-\omega_0-G \overline{ x}$ is the detuning between input and intracavity field renormalized by the presence of a power-dependent shift, and $\ucav{}$ represents the intracavity energy.

\begin{figure}
    \centering
    \includegraphics[width=0.45\textwidth]{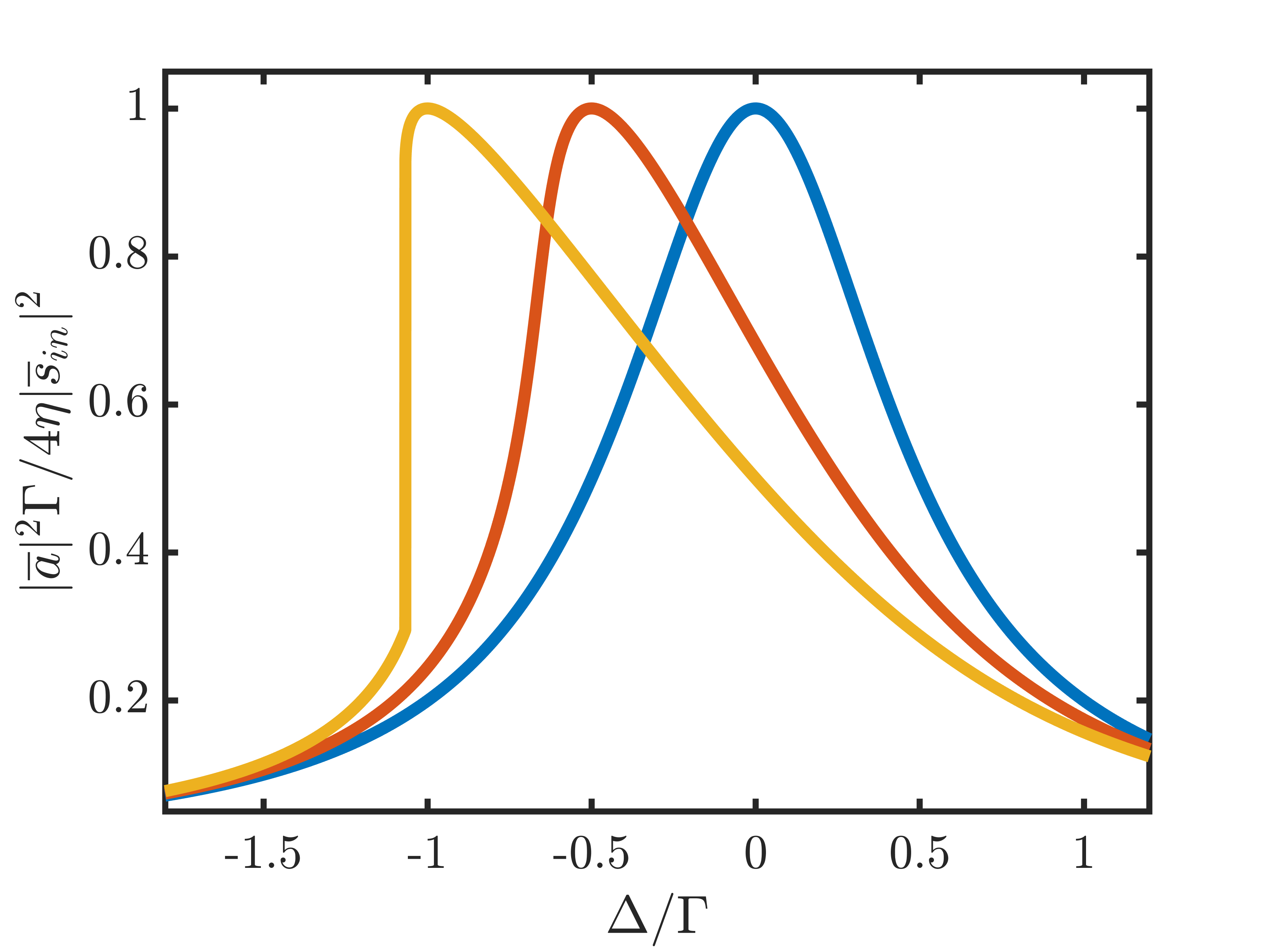}
    \caption{
    Normalized intracavity energy as a function of the detuning between the control and bare cavity frequencies, $\Delta=\omega_c-\omega_0$, in the presence of a constant driving field $\overline{s}_{in}$. 
    The increased coupled power (and,  consequently, resonance shift) at lower detuning results in a characteristic saw-tooth lineshape typical of bistable systems.
    The blue line represents the system response at low control power (peak cavity energy: $\ucav{}=10^{-3}\ucav{b}$), the red (peak $\ucav{}=\ucav{b}$) and yellow (peak $\ucav{}=2\ucav{b}$) curves represent the system response at increasing input power.
    } 
    \label{fig:fig2}
\end{figure}

By analogy with the theory of optical bistability for a localized mode coupled to a Kerr medium \cite{Soljacic2002}, Eqs.~\eqref{eq:steady_state} can be re-written in the form:

\begin{equation}
    \frac
    {\ucav{}}
    {\abs{\overline{s}_{in}}^2} =
    \frac
    {4\eta/\Gamma}
    {1+
    \left(2\Delta / \Gamma + \ucav{} /
    \ucav{b}
    \right)^2}
    \label{eq:bistability}
\end{equation}
\\
where we defined the characteristic bistability energy $ \ucav{b}=-{\gamma\Gamma}/{(2G \beta)}$. 
The solution of Eq.~\eqref{eq:bistability} may exhibit a characteristic ``sawtooth" line-shape as a function of the detuning $\Delta=\omega_c-\omega_0$, and depending on $\ucav{}$, as shown in Fig.~\ref{fig:fig2}, and as already reported \cite{Carmon2004}. 
In fact, in the presence of a sufficiently intense driving field ($\ucav{}\gg\ucav{b}$) and appropriate detuning conditions ($\Delta > \frac{\sqrt{3}}{2}\Gamma$) there are three possible solutions to the nonlinear equation, associated with two stable and one unstable equilibrium states, respectively {\cite{Lugiato1984,Carmon2004}}. In actual physical systems, the appropriate stable solution should be chosen according to the experimental conditions.

\subsection{Dynamic solution: induced absorption and gain}
\label{sec:dynamic_solution}

\subsubsection{Linearized dynamics}
We will now investigate the system response in the presence of an intense control field, in the form given in Eq.~\eqref{eq:control}, and a weak ``probe field" $\delta s_{in}(t)=s_pe^{i(\omega_c-\Omega)t}$, such that the overall driving term can be written as:
\begin{equation}
    s_{in}(t)= \overline{s}_{in}e^{i\omega_c t} + s_pe^{i(\omega_c-\Omega)t}
\end{equation}
\\
where $\abs{{s}_{p}}^2 \ll \abs{ \overline{s}_{in}}^2$. Notice that the input intensity displays an optical beating occurring at the control-probe detuning frequency, $\Omega$.

Under this assumption, we will model both the optical resonator and the first-order system responses as a steady state (i.e., mean field) value, adding on top of it a small time-dependent perturbation:
\begin{subequations}
\begin{align}
a(t)&= \overline{a}e^{i\omega_c t} + \delta a(t)  \\
 x(t)&= \overline{ x} +  \delta x(t)
\end{align}
\label{eq:linearized_variables}
\end{subequations}

The weak probe field can then be included in the model by linearizing the dynamical Eqs.~\eqref{eq:dynamics_x} and \eqref{eq:dynamics_field} at the equilibrium point, following a conventionally employed approach \cite{Weis2010}. By inserting the dynamical variables defined in  Eqs.~\eqref{eq:linearized_variables} into the coupled differential equations we get:

\begin{subequations}
\begin{align}
&\frac{d}{dt}\delta a(t) = 
\left( i \overline{\omega}_0 - \frac{\Gamma}{2}\right)\delta a(t)
+ i G \overline{a} \delta x(t)
+ \sqrt{\eta \Gamma} \delta s_{in} (t)
\\
& \frac{d}{dt}\delta x(t) = \beta \left( \overline{a}^* \delta a(t) + \overline{a}\delta a^*(t)\right) - \gamma\delta x(t) 
\end{align}
\label{eq:dynamics_lin}
\end{subequations}
\\
The solution can be found under the following ansatz:

\begin{align*}
\delta a(t) &= A_p^- e^{i(\omega_c-\Omega)t} + A_p^+ e^{i(\omega_c+\Omega)t} \\
 \delta x(t) &= X e ^ {-i\Omega t} + X^* e ^ {+i\Omega t} 
\end{align*}
\\
After including the latter into Eqs.~\eqref{eq:dynamics_lin} and separating terms according to their time dependence, we finally derive the following expressions for the oscillation amplitudes:

\begin{subequations}
\begin{align}
\label{eq:fields_ap_minus}
     A_p^- &= 
\frac{
iG\overline{a}X + \sqrt{\eta\Gamma}s_p
}{
i\left(\overline{\Delta}-\Omega\right) + \Gamma/2
}
\\ \label{eq:fields_ap_plus}
 A_p^+ &= 
\frac{
iG\overline{a}
}{
i\left(\overline{\Delta}+\Omega\right) + \Gamma/2
}\,X^*
\\ \label{eq:fields_temperature}
X&=\frac{\beta}{-i\Omega + \gamma} 
\left[\overline{a}^*A_p^- + \overline{a}(A_p^+)^* \right]
\end{align}
\label{eq:fields}
\end{subequations}

\subsubsection{Analytic solution}
\begin{figure}
    \centering
    \includegraphics[width=0.49\textwidth]{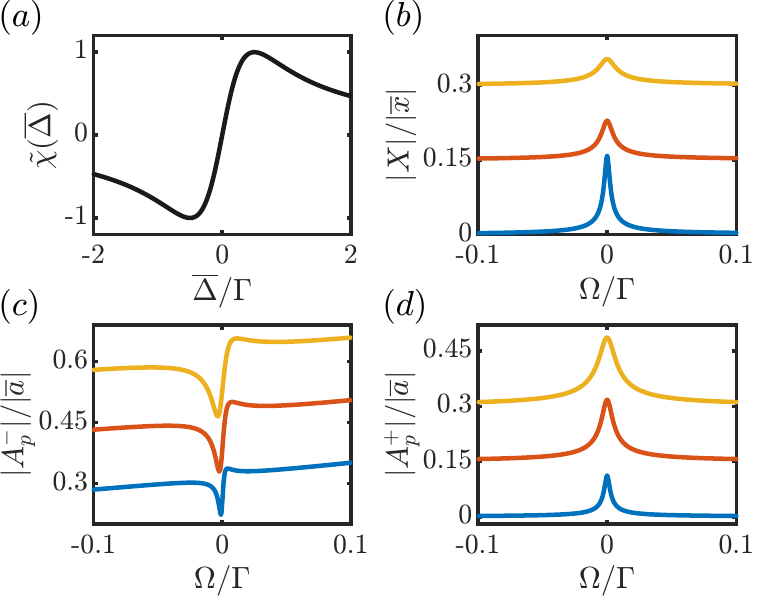}
    \caption{(a) Susceptibility of the induced transparency as a function of the  detuning between the control field and the shifted resonator mode frequencies. 
    (b-d) Oscillation amplitude/Stokes and anti-Stokes sidebands amplitude as a function of the probe-control detuning frequency, $\Omega$.
    The calculations were performed with system parameters $\gamma=10^{-3}\Gamma, G=-10^{-1}\Gamma,\beta=10^{-2}\Gamma$, and driving parameters $\abs{s_p}^2=0.1\abs{\overline{s}_{in}}^2$, $\overline{\Delta}=+\Gamma/2$ and $\ucav{}=(\ucav{b},3\ucav{b},5\ucav{b})$ from blue to yellow curve, respectively.
    All curves are spaced by a vertical offset of 0.15 for clarity.}
    \label{fig:fig3}
\end{figure}

The formal expressions given in Eqs.~\eqref{eq:fields} can now be used to derive  explicit analytic expressions for $A_p^-$, $A_p^+$ and $X$. In particular:

\begin{equation}
\boxed{
    X=\frac{\beta}
    {-i\Omega + \Gamma_{IT}}
    \cdot
    \frac{
    \overline{a}^* \sqrt{\eta\Gamma} 
    }{
    i\left(\overline{\Delta}-\Omega\right) + \Gamma/2
    }\cdot
    s_p
    \label{eq:temperature}
    }
\end{equation}
Here we defined an ``induced transparency linewidth'', whose physical meaning will become clear in the next subsection, in particular after  Eq.~\eqref{eq:beat_analytic}:
\begin{equation}
    \Gamma_{IT}=
    \gamma\left(
    1 + 
    \frac{\ucav{}}{\ucav{b}}\tilde{\chi}\left(\overline{\Delta}\right)
    \right)
    \label{eq:gamma_toit}
\end{equation}
where $ \ucav{b}$ represents the characteristic energy for optical bistability, already introduced in Sec.~\ref{sec:bistability}, while the parameter 
\begin{equation}
    \tilde{\chi}\left(\overline{\Delta}\right)\approx\frac{4\overline{\Delta}/\Gamma}{4\overline{\Delta}^2/\Gamma^2+1}
    \label{eq:IT_suscept}
\end{equation}
can be interpreted as an effective susceptibility for this nonlinear phenomenology \footnote{Notice that the full expression for the susceptibility is, in general, complex-valued: $\tilde{\chi}\left(\overline{\Delta}\right)
=  \frac{4\overline{\Delta}/\Gamma}
{4\left(\overline{\Delta}^2-\Omega^2\right)/\Gamma^2 -i4\Omega/\Gamma+1}$. The approximation assumes $\abs{\Omega}\ll\Gamma/2$.}, as it clearly appears from its plot in Fig.~\ref{fig:fig3}a.

The physical meaning of Eq.~\eqref{eq:temperature} can be summarized as follows. The term on the right-hand side of the product represents the optical response of the cavity under external excitation, in the absence of the induced transparency effect. 
This is described by a Lorentzian lineshape, centered at $\Omega=\overline{\Delta}$ and with full-width at half-maximum $\Gamma$. 
Conversely, the term on the left-hand side of the product is associated to the first-order response of the physical system. It also consists in a Lorentzian function, centered in $\Omega=0$ and with half-width at half-maximum $\Gamma_{IT}$. 

Thus, the oscillation amplitude $X$ (Fig.~\ref{fig:fig3}b) can be appreciably different from zero only if the control field is tuned in frequency within the optical resonator response ($\overline{\Delta}\sim\Gamma$) and the probe field is at the same time sufficiently close to the control frequency ($\Omega\sim\Gamma_{IT}$). 
In the case the first-order system response is much slower than the optical one ($\gamma\ll\Gamma$), as we will assume from now on, the overall spectral response of $X$ is dominated by the linewidth $\Gamma_{IT}$, and can be itself approximated by a Lorentzian curve. 
Notice that the width of the oscillation curve, $\Gamma_{IT}$ (i.e., the effective dissipation rate), depends on the control energy $\ucav{}$. 
In particular, it increases ($\Gamma_{IT}>\gamma$) in the blue-detuning regime ($\overline{\Delta}>0$), and it decreases ($0<\Gamma_{IT}<\gamma$) in the red-detuning regime ($\overline{\Delta}<0$). 
In both cases, for a given control energy $\ucav{}$, $\Gamma_{IT}$ is either maximized or minimized when $\overline{\Delta}=\pm\Gamma/2$. 
Thus, the effective dissipation rate of the first-order system is controlled by the action of the strong control field. 
This behavior is typical of EIT phenomenology, and it was also reported in connection with several EIT-analogues in the literature, such as \cite{Weis2010, Safavi-Naeini2011}.

Given the expressions \eqref{eq:temperature} and \eqref{eq:gamma_toit}, the anti-Stokes $A_p^+$ field can be analytically derived from Eq.~\eqref{eq:fields_ap_plus}:
\begin{equation}
    A_p^+\approx\frac{1}
    {i\Omega + \Gamma_{IT}}
    \cdot
    \frac{
    iG\beta\overline{a}^2 \sqrt{\eta\Gamma} 
    }{
    \overline{\Delta}^2-\Omega^2 + \Gamma^2/4
    }\cdot
    s_p^*
    \label{eq:ap_plus}
\end{equation}
The overall spectral lineshape of $A_p^+$ is also a Lorentzian curve (see Fig.~\ref{fig:fig3}d), with half-width at half-maximum $\Gamma_{IT}$.
Briefly, the field amplitude $A_p^+$ can be interpreted as a sideband generated by the modulation of the control field inside the cavity by the oscillation $X$, and similar considerations as for the $X$ solution lineshape apply. 

A similar expression for the Stokes field amplitude, $A_p^-$, can be analytically derived from Eq.~\eqref{eq:fields_ap_minus}:

\begin{multline}
    A_p^-=\left(
    1+
    \frac{1}
    {-i\Omega + \Gamma_{IT}}
    \cdot
    \frac{
    iG\beta\ucav{}
    }{
    i\left(\overline{\Delta}-\Omega\right) + \Gamma/2
    }\right)\\
    \cdot
    \frac{\sqrt{\eta\Gamma}}{
    i\left(\overline{\Delta}-\Omega\right) + \Gamma/2
    }
    \cdot  s_p
    \label{eq:ap_minus}
\end{multline}

In this case, the coherent mixing between the input probe field and the side-band generated by the modulation of the control field produces an asymmetric Fano-type lineshape (see Fig.~\ref{fig:fig3}c).

\subsubsection{Output signal}
We will now employ the analytic expressions derived above to construct the output signal. In particular, within an input-output formalism the output field can be expressed as \cite{Aspelmeyer2014}:
\begin{equation}
    s_{out}(t)=-\sqrt{\eta\Gamma}a(t)
    \label{eq:output_field}
\end{equation}
in which one assumes that the out-coupling rate, $\Gamma_{out}=\Gamma_{in}=\eta\Gamma$, is essentially the same as the input, for the sake of simplicity, and as it is common in many experimental situations.
Being directly proportional to the intracavity field amplitude, the output field contains three frequency components, oscillating at the carrier ($\omega_c$), Stokes ($\omega_c-\Omega$) and anti-Stokes ($\omega_c+\Omega$) frequencies, respectively. 
Notice that all the components' amplitude and phase depend both on the control and probe fields.

Focusing our attention to potential technological applications (e.g. optical pulse delay), we will define an output signal based on the optical beating occurring on the output field intensity, i.e. $I(t)=\abs{s_{out}(t)}^2$, where the output field amplitude is given by Eq.~\eqref{eq:output_field}. 
This depends on all the frequency components involved, and it is explicitly given as:

\begin{align*}
    I(t) 
    &=\eta\Gamma 
    \abs{ \overline{a} +
    A_p^-e^{-i\Omega t} +
    A_p^+e^{+i\Omega t}
    }^2 \\
    &=
    2\eta\Gamma\Re\left\{
    \overline{a}^* A_p^- + \overline{a} A_p^{+*}  \right\} \cos\Omega t \\
    &+2\eta\Gamma\Im\left\{
    \overline{a}^* A_p^- + \overline{a} A_p^{+*}  \right\} \sin\Omega t
    +\dots 
    \label{eq:beat}
\end{align*}
where in the last two lines we considered only the components oscillating at frequency $\Omega$. As aforementioned, the latter can be seen  as the beating frequency between control and probe fields. 
While an optical beating is observable also on the input field intensity, $\abs{s_{in}(t)}^2$, the amplitude and phase of this oscillation on $\abs{s_{out}(t)}^2$ depend exclusively on the underlying induced transparency phenomenology.

\begin{figure*}[t]
    \centering
    \includegraphics[width=0.75\textwidth]{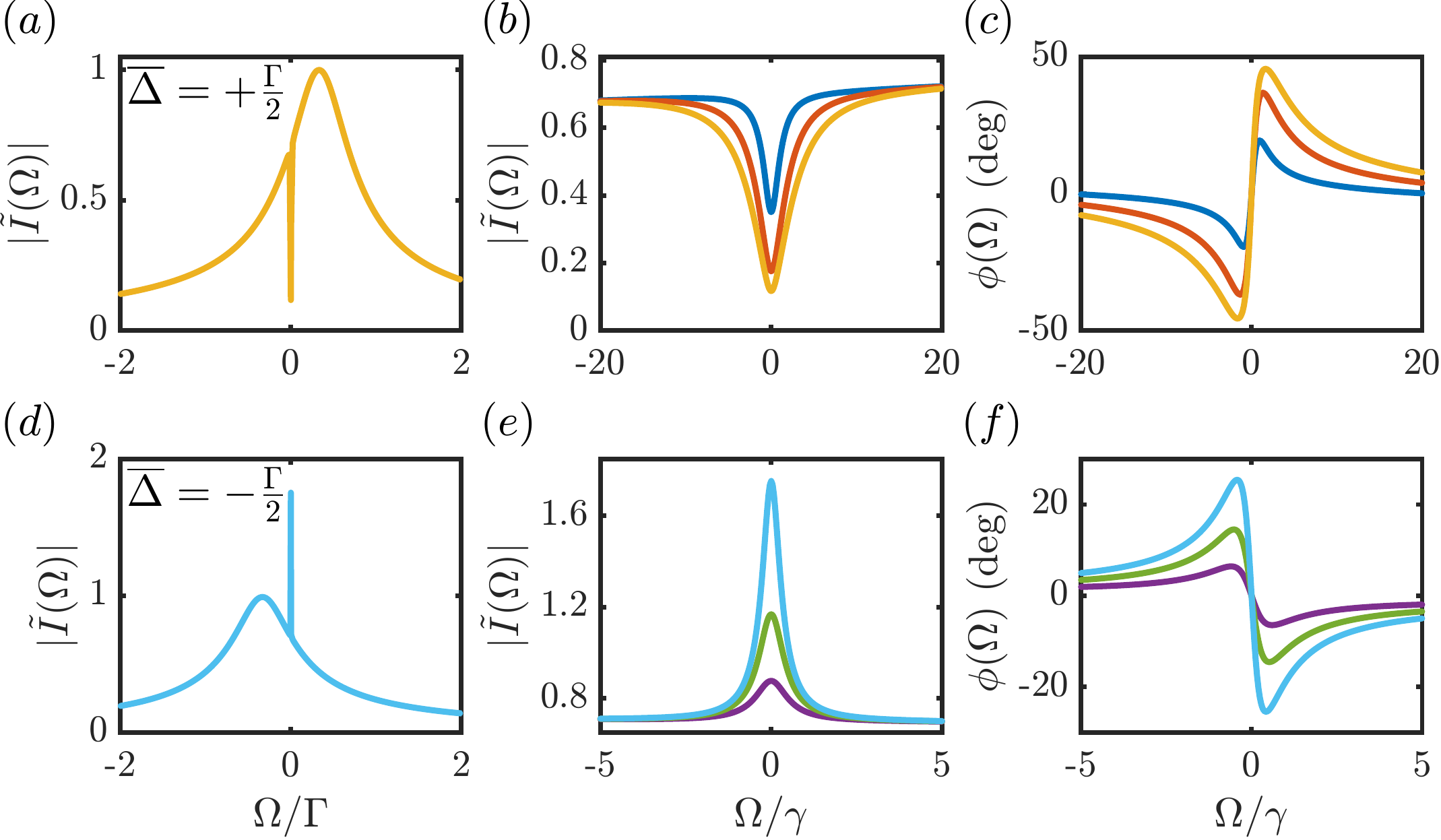}
    \caption{
    Output signal amplitude (normalized by the peak cavity response, at $\Omega=\overline{\Delta}$) and phase in blue- (a-c) and red- (d-f) detuning regimes, respectively. 
    The calculations assume the following physical parameters: $\gamma=10^{-3}\Gamma$, $G=-10^{-1}\Gamma$, $\beta=10^{-2}\Gamma$, and driving parameters $\overline{\Delta}=+\Gamma/2$, $\ucav{}=(\ucav{b},3\ucav{b},5\ucav{b})$ (a-c);  $\overline{\Delta}=-\Gamma/2$, $\ucav{}=(0.2\ucav{b},0.4\ucav{b},0.6\ucav{b})$ (d-f).
    }
    \label{fig:fig4}
\end{figure*}

In the frequency domain, the oscillation on the output intensity can be expressed in the complex form $\tilde{I}(\Omega)=2\eta\Gamma\left(\overline{a}^* A_p^- + \overline{a} A_p^{+*}\right)$. 
From comparison with Eqs.~\eqref{eq:fields_temperature} and \eqref{eq:temperature}, this is finally calculated as:
\begin{equation}
\boxed{
    \tilde{I}(\Omega)=\left(
    1-\frac{\Gamma_{IT}-\gamma}
    {-i\Omega + \Gamma_{IT}}\right)
    \cdot
    \frac{
    2\overline{a}^* \left(\eta\Gamma\right)^{3/2}
    }{
    i\left(\overline{\Delta}-\Omega\right) + \Gamma/2
    }\cdot
    s_p
    \label{eq:beat_analytic}
    }
\end{equation}
In the blue- (red-) detuning regime, the multiplying term within brackets represents a Lorentzian dip (peak) of width $\Gamma_{IT}$ and visibility $\abs{\mathcal{V}}$, respectively, such that:
\begin{equation}
\mathcal{V}=1-
{\frac{\gamma}{\Gamma_{IT}}}
\label{eq:visibility_beat}
\end{equation}
\\
where we have defined the visibility as the height of the dip (peak) spectral feature divided by the value of the unperturbed cavity response.

The amplitude of the output signal, Eq.~\eqref{eq:beat_analytic}, is explicitly shown in Fig.~\ref{fig:fig4}, 
for different driving and detuning conditions. As it is evidenced, the typical system response displays very narrow absorption or gain features within the broader Lorentzian response of the shifted resonator mode, similarly to the phenomenology already observed for optomechanically coupled oscillators \cite{Safavi-Naeini2011}.

\subsubsection{Phase response and group delay}
The visibility of the induced absorption or gain phenomenon is directly related to the already defined linewidth $\Gamma_{IT}$. 
As we shall see, this has a crucial role in the determination of the phase response, and hence the associated group delay or advance.

First, the phase associated to Eq.~\eqref{eq:beat_analytic} can be expressed as:
\begin{equation}
\phi(\Omega)=-\arg{\{\tilde{I}(\Omega)\} }
\approx\arctan\left\{
    \frac{\Omega\left(\Gamma_{IT}-\gamma\right)}
    {\Omega^2+\gamma\Gamma_{IT}}
    \right\}
\label{eq:phase_beat}
\end{equation}
\\
where we neglected the bare resonator phase response. Note that the above definition assumes an optical beating in the form $I(t)=|\tilde{I}|\cos\left(\Omega t + \phi\right)$. 
The above function has absolute maximum and minimum in $\Omega=\pm\sqrt{\gamma\Gamma_{IT}}$, which corresponds to a phase shift:
\begin{equation}
\phi^{peak}=\pm\arctan\left\{
    \frac{1}{2}\frac{\mathcal{V}}{\sqrt{1-\mathcal{V}}}
    \right\}
    \label{eq:peak_phase}
\end{equation}
\\
The effect of this dispersive system response on an optical pulse can be qualitatively understood as a phase shift, i.e., given by Eq.~\eqref{eq:phase_beat}, acting on all its frequency components.
In the region where $\abs{\Omega}<\sqrt{\gamma\Gamma_{IT}}$, the phase response is approximately linear, and the overall effect on the optical pulse is a time delay or advance.
This group delay is hereby defined as:
\begin{equation}
\tau_g(\Omega)=
-\frac{d\phi}{d\Omega}
\stackrel{\Omega\rightarrow0}{\approx}
-\frac{\mathcal{V}}{\gamma} \, ,
\label{eq:delay_beat}
\end{equation}
\\
which absolute value is clearly maximized for $\Omega \sim 0$. 
Notice that in the blue-detuning regime the delay is negative (group advance), and it asymptotically reaches the value $\tau_g^{min}=-1/\gamma$ for unit visibility. 
Conversely, in the red-detuning regime the delay is positive, and it diverges for $\Gamma_{IT} \rightarrow 0$.
The calculated phase response for different driving and detuning conditions is shown in Figs.~\ref{fig:fig4}c and~\ref{fig:fig4}f, and its phenomenology is consistent with the one commonly observed in bulk fast- and slow-light media \cite{Boyd2009}. 

\begin{figure}
    \centering
    \includegraphics[width=0.45\textwidth]{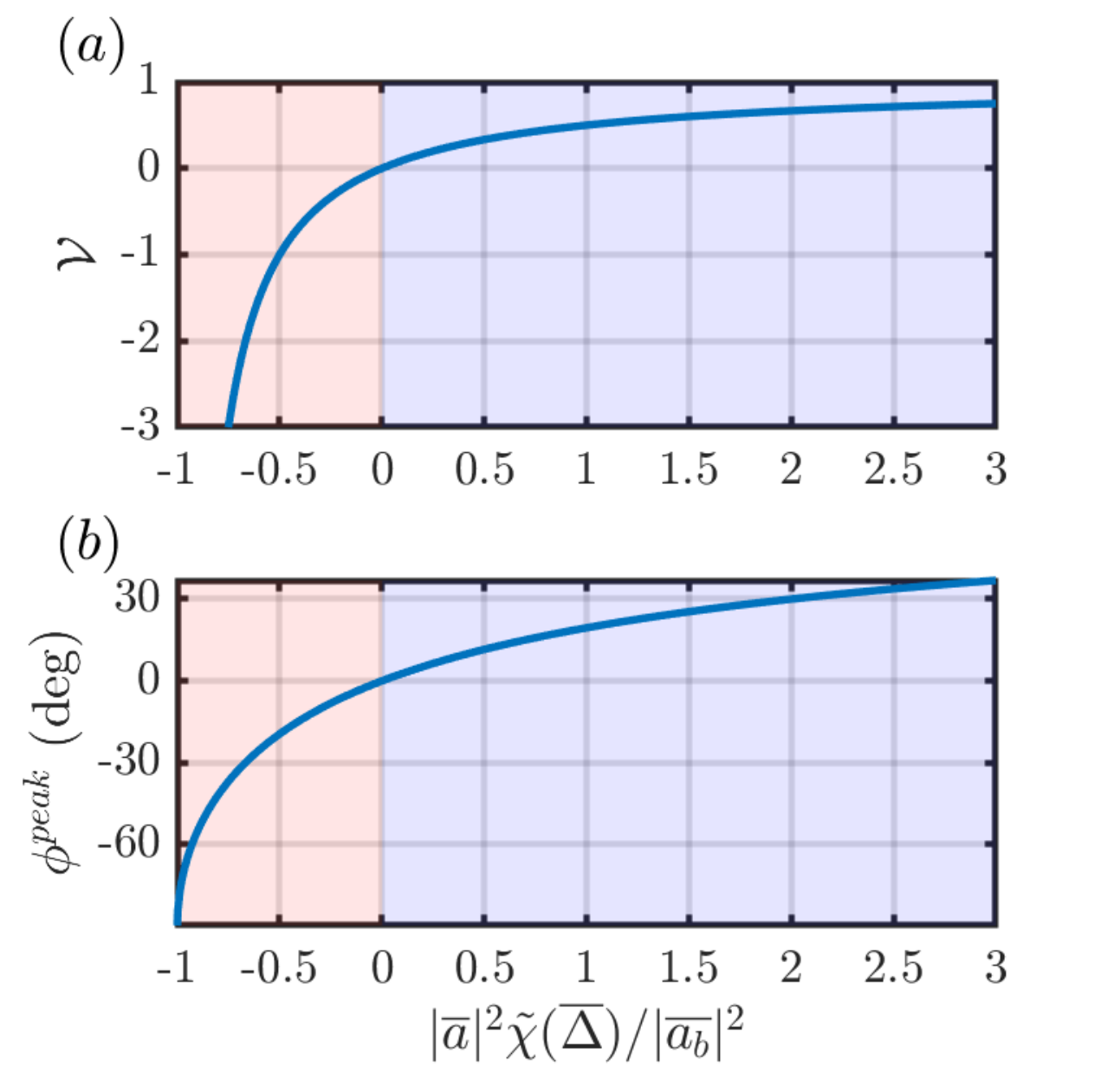}
    \caption{
     (a) Calculated trend for dip visibility as a function of the $\ucav{}\tilde{\chi}$ product, which value is determined by the control field intensity and detuning conditions. Negative values indicate a peak spectral feature.
    (b) Calculated trend for peak phase shift, Eq.~\eqref{eq:peak_phase}. The $+$ sign, associated to the local maximum/minimum at $\Omega=+\sqrt{\gamma\Gamma_{IT}}$, was chosen.
    Red (blue) shaded regions are associated to the respective detuning regimes. 
    }
    \label{fig:fig5}
\end{figure}

The visibility, $\mathcal{V}$, and the associated peak phase shift, $\phi^{peak}$, as defined respectively by Eqs.~\eqref{eq:visibility_beat} and \eqref{eq:peak_phase}, are plotted in Fig.~\ref{fig:fig5} as a function of the cavity energy $\ucav{}$ and detuning parameter $\tilde{\chi}$. The red- and blue-detuning regimes are explicitly indicated in the panels, respectively.

\section{Application to semiconductor microcavities}
\label{sec:applications}

The model and the solutions discussed so far are general, the only requirement being the coupling of a single-mode, Kerr-type nonlinear oscillator to a dissipative first-order system. We have shown that induced absorption and gain phenomena can easily be evidenced in such systems, as well as the associated group advance or delay. 

We will now discuss the actual relevance of these results in practical realizations of this model. In particular, we consider conventional semiconductor microcavities in standard material platforms, such as silicon or III-V alloys, which are typically characterized by intensity-dependent effective nonlinearities, such as thermo-optic or free-carrier dispersion effects \cite{Borghi2017}. 

\subsection{Thermo-optically induced transparency (TOIT)}
\label{sec:toit}

\subsubsection{Induced absorption and gain from thermo-optic nonlinearity}

The thermo-optic (TO) nonlinear effect consists in a local variation of the refractive index due to the heating of an optical medium. 
In a semiconductor microcavity, a variation of the refractive index in the region where the field is localized translates into a variation of the resonance frequency. 
According to first-order perturbation theory, the resonance shift is given by \cite{Carmon2004}:
\begin{equation}
    \frac{\Delta\omega_0}
    {\omega_0}=
    -\frac{\Delta n}{n}=
    -\frac{1}{n}\frac{dn}{dT}\Delta T
    =\alpha\Delta T
    \label{eq:to_effect}
\end{equation}
\\
where for simplicity we assumed the temperature variation $\Delta T$ (with respect to room temperature) to be uniform over the whole cavity region, and the field to be completely confined within the optical medium subject to TO effect. 
Moreover, we are neglecting any other temperature-related effect (e.g., thermal expansion) which may induce, by any extrinsic mechanisms, a further shift of the resonance wavelength. 
While these mechanisms are usually present in real-world microresonators, their impact can be quantified either from first principles \cite{Barclay2005} or by experimental measurement \cite{Debnath2017} of the resonance shift. 
The overall effect can be modelled, in the limit of a small temperature variation, by a correction to the TO coefficient, $\alpha$.

In the case in which linear absorption is the origin of the heating, and the optical intensity is time-varying, the temperature variation is also time-dependent. Its dynamics is then governed by the following rate equation \cite{Carmon2004}:

\begin{equation}
    C_p \frac{d}{dt}\Delta T(t)=
    \Gamma_{abs}\abs{a(t)}^2 - K \Delta T (t)
    \label{eq:th_decay_1t}
\end{equation}
\\
where $C_p$ is an effective heat capacity, $\Gamma_{abs}=\eta_{abs}\Gamma$ is the optical absorption rate, and $K$ is an effective heat conductance.

With these definitions, it is straightforward to recast Eqs.~\eqref{eq:dynamics_x} and \eqref{eq:dynamics_field} to appropriately describe the thermo-optic dynamics. 
This can be accomplished by formally identifying $x(t)=\Delta T(t)$, $\beta=\Gamma_{abs}/C_p$, $\gamma=K/C_p$, and $G=\omega_0\alpha$.
As a consequence, all of the results discussed in Sec.~\ref{sec:general_model} can be applied to single-mode optical resonators subject to linear absorption and TO nonlinear shift of its resonance frequency.

While thermal bistability is a well-known phenomenon in the domain of optical microcavites \cite{Carmon2004,Haret2009}, thermo-optically induced transparency (TOIT) has been demonstrated only very recently \cite{Clementi2020, Ma_SciAdv_2020}, and it can be regarded as a peculiar example of induced transparency from a first-order dynamical system obtained through mechanisms discussed in the previous Section.
Within this context, the previously introduced characteristic bistability energy is given by:
\begin{equation}
    \ucav{b}=
    -\frac{\gamma\Gamma}
    {2G\beta}
    =
    -\frac{K}
    {2Q\alpha\eta_{abs}}
\end{equation}
\\
where $Q=\omega_0/\Gamma$ is the quality factor of the bare cavity resonance, and $\eta_{abs}=\Gamma_{abs}/\Gamma$ is the absorption efficiency, i.e. the fraction of optical loss associated to light absorption. 
Notice that the formal expression of the bistability energy above does not depend on any dynamical parameter (such as $\gamma$), as it describes the static behavior of the resonator in the presence of a  control field with constant amplitude.

The value of $\ucav{b}$ can be very small for integrated resonators, with values of less than \SI{1}{\femto\joule} for silicon photonic crystal (PhC) cavities \cite{Haret2009} and few \si{\femto\joule} in microring resonators \cite{Almeida2004}, to name some of the most widespread microcavity designs.
Nevertheless, this nonlinearity has already been explored in the past for applications such as low-power all-optical switching \cite{Notomi2005}.
In the framework of TOIT, this translates in a very low threshold for the activation of the observed phenomenology \cite{Clementi2020}. 

The dynamical properties of the induced absorption or gain are governed by the thermal decay rate, $\gamma$.
As Eq.~\eqref{eq:gamma_toit} suggests, the already defined induced transparency linewidth, $\Gamma_{IT}$, will be (in most practical cases) quite close to $\gamma$ as an order of magnitude.
This value is, in general, larger for smaller sized microcavities, {as highlighted by recent experiments. In particular, a linewidth in the sub-\si{\mega\hertz} range has been reported in the case of TOIT measured in an integrated silicon PhC cavity \cite{Clementi2020}, and a sensibly slower dynamics in the case of a bulk Fabry-Pérot system, with a linewidth of the order of \SI{100}{\hertz} \cite{Ma_SciAdv_2020}.}
{Indeed, typical orders of magnitude for integrated resonators are} in the range $\gamma/2\pi\sim\SI{1}{\mega\hertz}$ for both silicon PhC cavities \cite{Tanabe2007,Haret2009} and microring resonators \cite{Priem2005,Pernice2010}, respectively.

\begin{figure}[t]
    \centering
    \includegraphics[width=0.3\textwidth]{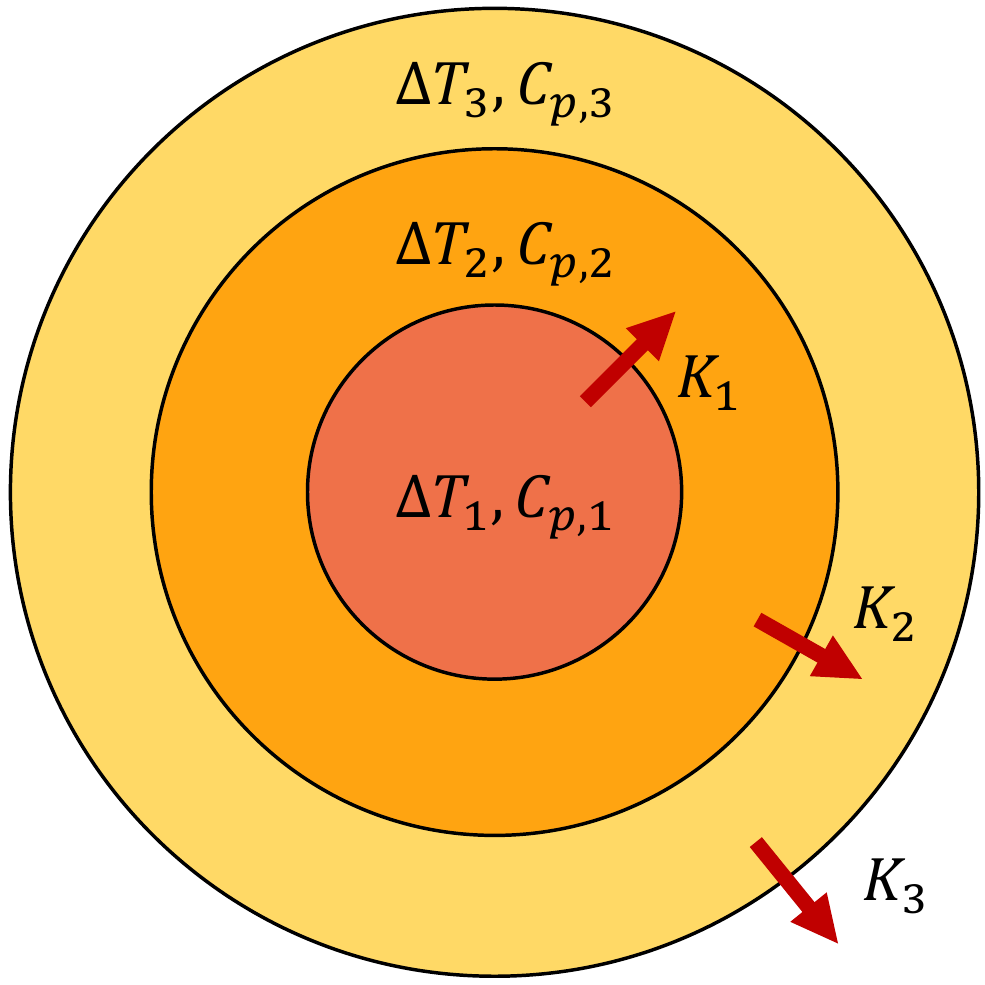}
    \caption{Discretized heat diffusion model. Each shell is characterized by a heat capacity $C_{p,i}$, and it exhibits a heat diffusion constant $K_i$ and a temperature offset $\Delta T_i$ with respect to the next outer shell.}
    \label{fig:heat_diffusion}
\end{figure}

\subsubsection{Modelling thermal diffusion}

The model for the TOIT effect detailed in the previous paragraph describes the thermal dynamics by means of a single temperature offset, $\Delta T(t)$, associated to a discrete thermal decay rate, $\gamma=K/C_p$, by means of the first-order differential equation~\eqref{eq:th_decay_1t}.
While this discretization allows to find a direct correspondence with the general case, i.e., Eqs.~\eqref{eq:dynamics_x} and \eqref{eq:dynamics_field}, it is usually quite of a rough approximation for the modeling of actual system dynamics.
Here, to complete this Section, we provide a generalization of TOIT based on a more accurate model of heat diffusion within a microstructure.

In the most general description, temperature is an intensive time-dependent quantity governed by the heat diffusion equation:
\begin{equation}
\rho c_p \frac{\partial}{\partial t} T(\mathbf{r},t) + \nabla \cdot \left(-\kappa \nabla T(\mathbf{r},t)\right) = \frac{\partial}{\partial t}u(\mathbf{r},t)
\label{eq:diffusion}
\end{equation}
\\
where $\rho$ is the medium density, $c_p$ is the mass specific heat, $\kappa$ is the thermal conductivity, and $\frac{\partial u}{\partial t}$ is a source term describing the heat flux density towards the system, in this case associated to the absorbed optical power.
In this context, Eq.~\eqref{eq:th_decay_1t} represents a discretized version of the heat diffusion equation, where the temperature offset with respect to room temperature is considered uniform over a finite volume region, i.e., $\Delta T(\mathbf{r},t) \rightarrow \Delta T(t)$. In this ``lumped elements" model, the intensive physical quantities, $c_p$ and $\kappa$, appearing in Eq.~\eqref{eq:diffusion} are respectively replaced by extensive ones, namely $C_p$ and $K$.

Following an approach already suggested in previous works \cite{Ilchenko1992,Carmon2004,Iadanza2020}, a more accurate description of the problem can be formulated by slicing the thermal distribution into $n$ concentric regions, whose definition depends on the detailed geometrical aspects of the system. A schematic representation is reported in  Fig.~\ref{fig:heat_diffusion}, for completeness.
Each region is labeled from 1 to $n$, such that $i=1$ is the innermost thermal shell, where the field is mainly confined, and $i=n$ is the outermost one.
The $i$-th region is characterized by a heat capacity $C_{p,i}$, an effective temperature offset $\Delta T_i$ with respect to the $(i+1)$-th one, and a heat diffusion constant $K_i$ towards it. Given these definitions, Eqs.~\eqref{eq:dynamics_x} and \eqref{eq:dynamics_field} are straightforwardly generalized as follows:
\begin{subequations}
\begin{align}
&\frac{d}{dt}a(t) =
\left(i\omega_0 - \frac{\Gamma}{2} \right)a(t) + 
i G \Delta T(t) a(t) +
\sqrt{\eta \Gamma} s_{in}(t)
\\
&\frac{d}{dt}\Delta T_1(t)=
\beta \abs{a(t)}^2 - 
\gamma_{1,1} \Delta T_1 (t)
\\
&\frac{d}{dt}\Delta T_2(t)=
\gamma_{1,2} \Delta T_1 (t) - 
\gamma_{2,2} \Delta T_2 (t)
\\
&\dots \notag
\\
&\frac{d}{dt}\Delta T_n(t)=
\gamma_{n-1,n} \Delta T_{n-1} (t) - 
\gamma_{n,n} \Delta T_n (t)
\end{align}
\label{eq:dynamics_multi}
\end{subequations}
\\
where the overall temperature offset experienced by the field is $\Delta T(t) =\sum_{i=1}^n \Delta T_i(t)$, while the rates connecting different regions are defined $\gamma_{i,j}=K_i/C_{p,j}$ and $\beta=\Gamma_{abs}/C_{p,1}$. With these definitions and the assumptions made in the previous paragraph, 
Eqs.~\eqref{eq:dynamics_multi} describe with arbitrarily high accuracy any thermal diffusion process that is consistent with the symmetry and geometrical details of the problem.

In the presence of a monochromatic control field, a solution formally similar to Eq.~\eqref{eq:steady_state} can be derived:
\begin{subequations}
\begin{align}
&\overline{a} = 
\frac{\sqrt{\eta \Gamma}}
{i\left(\omega-\omega_0-G\overline{\Delta T}\right) + \Gamma/2}
\overline{s}_{in} \\
&\overline{\Delta T} = \frac{\Gamma_{abs}}{K} \abs{\overline{a}}^2
\end{align}
\label{eq:steady_state_multi}
\end{subequations}
\\
where $\overline{\Delta T} =\sum_{i=1}^n \overline{\Delta T}_i$ is the overall temperature offset, and $K=\left(\sum_{i=1}^nK_i^{-1}\right)^{-1}$ is the total series conductance.
With these definitions, the steady state result obtained from the generalized model is formally identical to the one obtained in the simplified one, nicely.
This result justifies \textit{a posteriori} the simplified description in the steady state condition, which can be effectively described by a single temperature offset, $\Delta T$, and a single heat conductance, $K$.

\begin{figure}[t]
    \centering
    \includegraphics[width=0.49\textwidth]{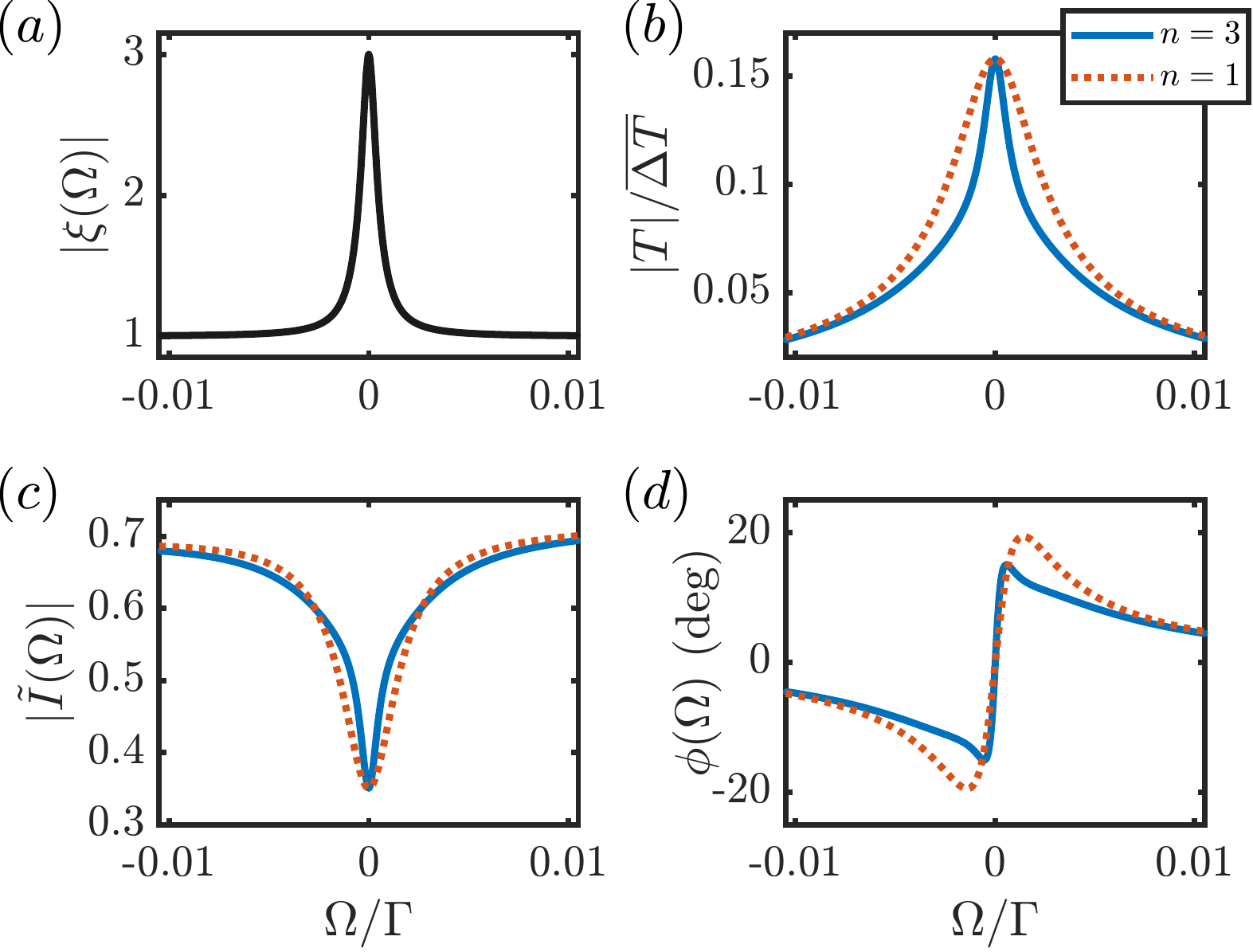}
    \caption{
    Comparison between the TOIT response for the generalized (blue solid line) and the simplified (red dashed line) models, respectively.
    (a) Thermal response function $\xi(\Omega)$ as a function of the probe-control detuning.
    (b) 
    Temperature oscillation amplitude.
    (c)
    Amplitude and 
    (d)
    phase response for the output signal.
    The calculation assumes the following parameters: $G=-10^{-1}\Gamma$, $\beta=10^{-2}\Gamma$, $\overline{\Delta}=+\Gamma/2$ and $\ucav{}=\ucav{b}$.
    For the non-refined model ($n=1$) we chose $\gamma=10^{-3}\Gamma$, while for the 
    refined model ($n=3$) we chose $\gamma_{1,1}=3\times10^{-3}\Gamma$, $\gamma_{1,2}=\gamma_{2,2}=\gamma_{1,1}/4$, $\gamma_{2,3}=\gamma_{3,3}=\gamma_{1,1}/8$. Note that by this choice of parameters, the static temperature offset $\overline{\Delta T}$ is the same in both scenarios.
    }
    \label{fig:fig7}
\end{figure}

Following the steps discussed in App.~\ref{app:derivation}, it is possible to derive an analytic expression for the temperature oscillation amplitude, $T(\Omega)$, and the output signal, $\tilde{I}(\Omega)$, in the framework of this generalized model:
\begin{equation}
    T(\Omega)=\xi(\Omega)\frac{\beta}
    {-i\Omega + \Gamma_{IT}(\Omega)}
    \cdot
    \frac{
    \overline{a}^* \sqrt{\eta\Gamma} 
    }{
    i\left(\overline{\Delta}-\Omega\right) + \Gamma/2
    }\cdot
    s_p
    \label{eq:temperature_multi}
\end{equation}
\begin{equation}
    \tilde{I}(\Omega)=\left(
    1-\frac{\Gamma_{IT}(\Omega)-\gamma_{1,1}}
    {-i\Omega + \Gamma_{IT}(\Omega)}\right)
    \cdot
    \frac{
    2\overline{a}^* \left(\eta\Gamma\right)^{3/2} 
    }{
    i\left(\overline{\Delta}-\Omega\right) + \Gamma/2
    }\cdot
    s_p
    \label{eq:beat_analytic_multi}
\end{equation}
\\
in which:
\begin{equation}
    \Gamma_{IT}(\Omega)=
    \gamma_{1,1}\left(
    1 + 
    \frac{\abs{\overline{a}}^2}
    {\abs{\overline{a}_b}^2}\xi'(\Omega) \tilde{\chi}\left(\overline{\Delta}\right)
    \right)
    \label{eq:gamma_toit_multi}
\end{equation}
\\
{where $\xi' (\Omega) = K \xi (\Omega) /K_1$, and:
\begin{equation}
    \xi(\Omega)=
    1 + \sum_{j=2}^{n}\left( \prod_{i=2}^j \frac{\gamma_{i-1,i}}{-i\Omega + \gamma_{i,i}} \right) \, .
\end{equation}
\\
The latter expression clearly displays a recursive hierarchical structure.}
Notice that $\Gamma_{IT}(\Omega)$ is now a frequency-dependent quantity, and it cannot be simply interpreted as a ``linewidth'', as done before in connection with Eq.~\eqref{eq:gamma_toit}. 
However, this quantity still provides relevant information for the description of the induced transparency visibility. 
In fact, the thermal response function, $\xi(\Omega)$ (plotted in Fig.~\ref{fig:fig7}), and its normalized form, $\xi'(\Omega)$, contain all the information about the thermal process dynamics, quantified by the decay rates $\gamma_{i,j}$.
A comparison between this result (by assuming $n=3$) and the simplified model with a single thermal decay time is explicitly shown in Fig.~\ref{fig:fig7}. As it can be noticed, the simplified model is a reasonably good approximation for the quantitative estimation of the zero-detuning response, but it fails to correctly capture the detailed lineshapes, which may significantly deviate from a single Lorentzian.

Similar figures of merit as for the general model can be derived. In particular, the expression for the  visibility is basically the same as the one previously introduced in Eq.~\eqref{eq:visibility_beat}, and it reads:
\begin{equation}
    \mathcal{V}=1-\frac{\gamma_{1,1}}{\Gamma_{IT}(0)}
\end{equation}
\\
In contrast, the expressions for the phase and the group delay become more complicated in this case than the ones introduced in Eqs.~\eqref{eq:phase_beat} and Eq.~\eqref{eq:delay_beat}, respectively. 
A full derivation is detailed in Appendix~\ref{app:derivation}, for completeness. 
In particular, for small values of $\Omega$  it is found that the group delay as a function of $\mathcal{V}$ shows a trend similar to the one given by Eq.~\eqref{eq:delay_beat}:
\begin{equation}
    \tau_g(\Omega\rightarrow0) \approx
    -\frac{\mathcal{V}}{\gamma_{eff}}
    \label{eq:delay_beat_multi}
\end{equation}
\\
where $\gamma_{eff}$ is an effective thermal decay rate, which is a function of the $\gamma_{i,j}$ parameters. 
The minimum (negative) group delay achieved asymptotically in the blue-detuning regime is then given by $\tau_g^{min}=-1/\gamma_{eff}$.
As an example, for the case $n=3$ this is given by:

\begin{equation*}
    \tau_g^{min}=
    -\frac
    {\gamma_{1,2}/\gamma_{2,2}^2 +
    \left(\gamma_{2,2}+\gamma_{3,3}\right)
    \gamma_{1,2}\gamma_{2,3}/
    \gamma_{2,2}^2\gamma_{3,3}^2}
    {1+\gamma_{1,2}/\gamma_{2,2}+
    \gamma_{1,2}\gamma_{2,3}/\gamma_{2,2}\gamma_{3,3}}
\end{equation*}


\subsection{Free-carrier induced transparency (FCIT)}
\label{sec:fcit}

Free-carrier dispersion (FCD) is an ubiquitous phenomenon in semiconductor microcavities, and its behavior has been extensively studied in silicon devices \cite{Soref1987,Borghi2017}.
It consists in a modification of the refractive index induced by the presence of hot carriers in the valence or conduction bands of a semiconductor material, which can be modeled by a simple Drude theory to a first approximation \footnote{
We notice that experimental results show that the Drude model is not sufficient to account for the whole phenomenology associated to FCD, in some practical cases \cite{Soref1987}.
In particular, in the case of silicon the contribution to dispersion from electrons and holes is slightly different:
\begin{equation*}
    \frac{\Delta n}{n}
    =-\frac{1}{n}
    \left(
    \zeta_e N_e + \zeta_h N_h^{0.8}
    \right)
\end{equation*}
For simplicity, we will ignore this correction in the following discussion.
}.

If a localized resonant mode is confined in a region affected by FCD, the associated resonance shift will be, to first-order \cite{Barclay2005}:
\begin{equation}
    \frac{\Delta\omega_0}
    {\omega_0}=
    -\frac{\Delta n}{n}=
    -\frac{1}{n}\frac{dn}{dN}N
    =\zeta N
    \label{eq:fcd}
\end{equation}
\\
where $N$ is the effective carrier density, which we assume to have a uniform distribution over the whole cavity region.
We also assume the field to be completely confined within the optical medium subject to FCD.
The FCD coefficient, $\zeta$, is typically positive, i.e., the presence of a carrier density yields a blue-shift of the cavity resonance.

If the population of free-carriers is the result of an excitation by the electromagnetic field associated to the localized mode, FCD phenomenologically behaves as an effective nonlinearity, similar to the TO effect but with opposite sign.
Specifically, if the free-carrier population is generated by linear absorption (e.g., absorption from defect states), the free-carrier dynamics is described by the following rate equation \cite{Iadanza2020}:
\begin{equation}
    \frac{d}{dt} N(t)=
    \frac{\Gamma_{abs}}{V\hbar\omega}\abs{a(t)}^2 - \frac{1}{\tau_c} N (t)
    \label{eq:fc_dynamics}
\end{equation}
\\
where $\omega$ is the electromagnetic field frequency, $\Gamma_{abs}$ is the absorption rate, $\tau_c$ is the free-carrier recombination lifetime, and $V$ is an effective volume occupied by the free-carriers.

Again, it is immediate to recast Eqs.~\eqref{eq:dynamics_x} and \eqref{eq:dynamics_field} to properly describe the FCD dynamics. 
Specifically, here we formally identify $x(t)=N(t)$, $\beta=\Gamma_{abs}/V\hbar\omega$, $\gamma=1/\tau_c$, and $G=\omega_0\zeta$.
As a consequence, the predicted phenomenology described in Section \ref{sec:general_model} is observable in semiconductor optical cavities affected by linear absorption and FCD.
In contrast to TOIT, this novel phenomenon, which we will refer as free-carrier induced transparency (FCIT), has not been reported experimentally to date, yet.
The resonance shift induced by FCD has been investigated in the past mainly in the framework of fast all-optical switching \cite{Nozaki2010,Belotti2010}.
In this context, the FCD mechanism typically consists in a much faster response time and a comparable characteristic bistability energy as compared to TO effect. Similar considerations then apply to FCIT as compared to TOIT.

In fact, the characteristic bistability energy is here given by:
\begin{equation}
    \ucav{b}=
    -\frac{\gamma\Gamma}
    {2G\beta}
    =
    -\frac{\hbar V}
    {2\tau_c\zeta\eta_{abs}}
\end{equation}
\\
where we assumed $\omega\approx\omega_0$.
Notice that, since the sign of interaction is opposite to the case of TO effect (here, $G>0$), FCD gives rise to a reversed sawtooth-shaped response in the steady state scenario as compared to the one already reported in Fig.~\ref{fig:fig2}.
In practical cases, a combination of TO and FCD nonlinearities may occur, which may manifest into the dynamical cavity response. In some circumstances, this gives rise to an unstable behavior such as self-pulsations, as already reported \cite{Pernice2010}.

In the case of FCIT, the most remarkable consequence of the reversed sign of the interaction is an opposite phenomenology when compared to the TOIT scenario: in the blue-detuning regime ($\overline{\Delta}>0$), a FCIT peak is expected, associated to a narrowing of the induced transparency linewidth ($\Gamma_{IT}<\gamma$). 
Conversely, a FCIT dip with associated  linewidth broadening ($\Gamma_{IT}>\gamma$) is expected in the red-detuning regime ($\overline{\Delta}<0$).
Apart from this difference, the phenomenology to be expected with these two nonlinear processes is essentially identical, and a generalized model accounting for the spatial carrier diffusion can be formulated with an approach similar to the one presented in the previous paragraph.
The formulation of such a model, based on the carrier diffusion equation rather than the heat diffusion law, is straightforward but goes beyond the scope of the present work.

In practical semiconductor devices, FCIT exhibits a higher characteristic bistability energy $\ucav{b}$ as compared to TOIT. 
The existing literature reports all-optical switching energy values ranging from $\sim\SI{100}{\femto\joule}$ to $\sim\SI{10}{\pico\joule}$ in silicon PhC cavities \cite{Tanabe2005a,Belotti2010}, and sub-\si{\femto\joule} values reported for III-V devices exploiting the band-filling dispersion (BFD) nonlinearity \cite{Nozaki2010}, an analogous but physically different dispersive mechanism to FCD. Switching energies in the order of \SI{1}{\pico\joule} have been reported for microring resonators \cite{Waldow2008}.

\subsection{Discussion and comparison}
\label{sec:disc}

We have shown two different examples of effective nonlinearities that could be treated within the unified framework introduced in the previous Section, yielding remarkably similar phenomenology associated with EIT physics. In particular, applications requiring slow and fast light may benefit from the engineering of such first-order dynamical systems.

In this respect, it is worth reminding that a fast response time is particularly beneficial for applications requiring a wide operational bandwidth, such as fast optical memories. 
However, the overall delay (or advance) achievable is limited, in general, by a key figure of merit such as the delay-bandwidth product, $\Delta t\Delta\omega\le 2\pi$. Hence, the narrower the bandwidth, the larger the delay time that could be achieved. 
Within this context, it is worth stressing that TOIT allows great flexibility offered at a design-level by the engineering of the resonator thermal properties \cite{Iadanza2020}. Another key feature of TOIT is the possibility to overcome the delay-bandwidth limit by cascading multiple resonators \cite{Morichetti2008}.
This solution is practically hindered in many EIT-analogues, due to the tight requirements of matching an optical and an atomic, mechanical or acoustic resonance, while it would be relatively easy to implement in TOIT-based devices.

Despite the higher threshold energy as compared to TOIT, FCIT has significant advantages in terms of bandwidth, which is determined by the free-carrier lifetime, $\tau_c$, rather than the thermal decay time.
Typical values can be of the order of \SI{100}{\pico\second} for PhC cavities \cite{Notomi2010,Borghi2017}, and of the order of \SI{1}{\nano\second} for whispering gallery mode resonators \cite{Osgood2009}. 
In both cases, $\tau_c$ can be further reduced by ion-implantation \cite{Tanabe2007,Waldow2008}.
The resulting induced transparency linewidth, and hence slow or fast light bandwidth, lies in the \si{\giga\hertz} range, opening a pathway to applications such as broadband delay lines and optical memories, even at room temperature and in cost-effective material platforms.
Finally, while it is worth noticing that the maximum delay or advance theoretically achievable with FCIT would be much lower than the TOIT case, also FCIT can be easily cascaded along multiple resonators, thus considerably extending the potential delay achievable above these limits.

\section{Summary and outlook}

We have presented a general model describing a physical system governed by first-order dynamics and coupled to a single-mode optical resonator, showing how its solution can lead to the observation of analogue EIT phenomenology in a driving configuration in which a weak probe beam is superimposed to a strong control field.
From a set of two coupled equations of motion, we have derived analytic expressions for the key physical quantities involved, both in the steady state and in a control and probe scenario.
For the latter case, we showed that induced transparency emerges as an interference effect between the cavity control and probe fields, mediated by the effective interaction provided by the first-order dynamical system.
This results in induced absorption (spectral hole) in the control-cavity blue-detuning regime ($\overline{\Delta}>0$), and induced gain or amplification (spectral anti-hole) in the red-detuning regime ($\overline{\Delta}<0$) for a Kerr-type effective interaction ($G<0$).
The former case is associated to fast light and the broadening of the induced transparency linewidth, while the latter case is associated to slow light and the narrowing of the induced transparency feature.
The relevant figures of merit, including the spectral linewidth $\Gamma_{IT}$, the visibility $\mathcal{V}$, and the group delay $\tau_g$ were explicitly derived, and linked to the properties of the physical system, such as the characteristic optical bistability energy $\ucav{b}$ and the first-order system damping rate $\gamma$.

In Sec.~\ref{sec:applications}, we discussed the specific correspondence of this general model to the physics of semiconductor microcavities subject to effective Kerr-type nonlinearities.
In particular, our focus was on two different dispersive mechanisms, namely the thermo-optic effect and the free-carrier dispersion, which are typical of several semiconductor photonic devices.
We evidenced how both effects can be driven by the absorption of the cavity field, and they can thus be regarded as effective Kerr-type nonlinearities, each described by a characteristic first-order differential equation.
The resulting induced transparency effects are theoretically predicted to yield a group advance or delay in the $\si{\micro\second}$ range for TOIT and in the $\si{\nano\second}$ range for FCIT, both hardly achieved on-chip by current photonic technology \cite{Alexoudi2020}. 
As a comparison, a $\si{\micro\second}$-scale delay achieved in an on-chip delay line would require a $\sim\SI{100}{\meter}$ long integrated waveguide, with the current silicon photonics technology.
Moreover, the two effects are associated with operational bandwidths in the $\si{\mega\hertz}$ and $\si{\giga\hertz}$ ranges, respectively, which could be further increased by cascading multiple resonators without the need to match any mechanical or acoustic resonance.
Furthermore, the transparency bandwidth could be controlled, in principle, by engineering the thermal or free-carrier properties of the microstructured resonators, for instance by heat diffusion design \cite{Iadanza2020} or ion-bombardment \cite{Tanabe2005a}. It is also worth stressing that TOIT has recently been observed in two different experimental scenarios \cite{Clementi2020,Ma_SciAdv_2020}, while FCIT has not been reported at time of writing, which we believe will stimulate new experimental research in this direction.

Finally, the general and unified framework presented in this work can be extended to other physical systems displaying a first-order response and coupled to a single-mode oscillator, opening a pathway towards the demonstration of novel induced transparency analogues and the achievement of slow light in yet unforeseen platforms, thus revealing novel viable approaches towards the realization of optical and quantum memories.

\begin{acknowledgments}
This work was partly supported by the EU H2020 QuantERA ERA-NET Co-fund in Quantum Technologies project CUSPIDOR, co-funded by the Italian Ministry of Education, University and Research (MIUR), and by MIUR: “Dipartimenti di Eccellenza Program (2018-2022)”, Department of Physics, University of Pavia. 
\end{acknowledgments}

\appendix
\section{Derivation for the refined TOIT model}
\label{app:derivation}
Following the steps already discussed in Sec.~\ref{sec:dynamic_solution}, we will now derive the linearized equations of motion in the hypothesis of a weak probe field. 
The procedure is formally identical to the previous one for what concerns the field amplitude, while for the effective temperature the following additional assumptions are made:
\begin{subequations}
\begin{align}
\Delta T_i(t)&= \overline{\Delta T}_i + \delta T_i(t)\\
\delta T_i(t) &= T_i e ^ {-i\Omega t} + T_i^* e ^ {+i\Omega t} 
\end{align}
\end{subequations}
\\
where:
\begin{equation*}
    \overline{\Delta T} =\sum_{i=1}^n \overline{\Delta T}_i \qquad
    \delta T =\sum_{i=1}^n \delta T_i \qquad
    T =\sum_{i=1}^n T_i
\end{equation*}
\\
The linearized equations of motion thus read:
\begin{subequations}
\begin{align}
&\frac{d}{dt}\delta a(t) = 
\left( i \overline{\omega}_0 - \frac{\Gamma}{2}\right)\delta a(t)
+ i G \overline{a} \delta T(t)
+ \sqrt{\eta \Gamma} \delta s_{in} (t)
\\
& \frac{d}{dt}\delta T_1(t) = \beta \left( \overline{a}^* \delta a(t) + \overline{a}\delta a^*(t)\right) - \gamma_{1,1}\delta T_1(t) \\
& \frac{d}{dt}\delta T_i(t) = 
\gamma_{i-1,i}\delta T_{i-1}(t) - 
\gamma_{i,i}\delta T_i(t)
\qquad (i>1)
\end{align}
\label{eq:dynamics_lin_multi}
\end{subequations}
\\
We notice that the solution is formally similar to the one already found for a single thermal rate, Eqs.~\eqref{eq:fields}, although the temperature oscillation $T$ now derives from all the temperature offsets $T_i$:

\begin{subequations}
\begin{align}
     A_p^- &= 
\frac{
iG\overline{a}T + \sqrt{\eta\Gamma} s_p
}{
i\left(\overline{\Delta}-\Omega\right) + \Gamma/2
}
\\ 
 A_p^+ &= 
\frac{
iG\overline{a}
}{
i\left(\overline{\Delta}+\Omega\right) + \Gamma/2
}\,T^*
\\
T_1&=\frac{\beta}{-i\Omega + \gamma_{1,1}} 
\left(\overline{a}^*A_p^- + \overline{a}(A_p^+)^* \right) \label{eq:fields_multi_T}
\\
T_i&=\frac{\gamma_{i-1,i}}{-i\Omega + \gamma_{i,i}}T_{i-1}
\qquad (i>1)
\end{align}
\label{eq:fields_multi}
\end{subequations}
\\
From the last equation:
\begin{equation}
    T_i=
    \frac{\gamma_{i-1,i}}{-i\Omega + \gamma_{i,i}}\cdot
    \frac{\gamma_{i-2,i-1}}{-i\Omega + \gamma_{i-1,i-1}}\cdot
    \dotso \cdot
    \frac{\gamma_{1,2}}{-i\Omega + \gamma_{2,2}}\cdot
    T_1
\end{equation}
\\
which is valid for any $i>1$. 
{We can find a general relation between $T$ and $T_1$:
\begin{equation}
\begin{split}
    T & = 
    \sum_{i=1}^n T_i =
    {\xi(\Omega)}
    T_1
\end{split}
\end{equation}
\\
In the last step we defined the thermal response function $\xi(\Omega)$ such that:
\begin{equation}
\begin{split}
   \xi(\Omega) & = 1 +
    \frac{\gamma_{1,2}}{-i\Omega + \gamma_{2,2}} + \\
    &+ \frac{\gamma_{1,2} \gamma_{2,3}}{(-i\Omega + \gamma_{2,2})(-i\Omega + \gamma_{3,3})} +
    \dots +
    \prod_{i=2}^n \frac{\gamma_{i-1,i}}{-i\Omega + \gamma_{i,i}} \, .
\end{split}
\end{equation}
\\
Notice that $\xi(\Omega)$ has an absolute maximum in $\Omega=0$, for which $\xi(0)=K_1/K$.}

Applying this relation to the field components \eqref{eq:fields_multi}, it is immediately possible to recover the formal expressions for the fields derived in the general model.
In particular, by directly comparing with Eq.~\eqref{eq:fields_multi_T}, we notice that all the expressions previously obtained, including $\beta$ and $\Gamma_{IT}$, can be generalized by substituting $\beta\rightarrow\xi(\Omega)\beta$ and $\gamma\rightarrow\gamma_{1,1}$.
The resulting expressions for $T(\Omega)$, $\tilde{I}(\Omega)$, and $\Gamma_{IT}(\Omega)$ are reported in the main text.

In calculating the phase response associated to the output signal, we first subdivide $\xi(\Omega)$ in a real and imaginary part:
\begin{gather*}
    \xi(\Omega)=\xi^r+i\xi^i
    \\
    \Gamma_{IT}(\Omega) = \gamma_{1,1} - \xi(\Omega)\zeta
\end{gather*}
We then express:
\begin{equation*}
    \tilde{I}(\Omega)=
    \frac{-i\Omega+\gamma_{1,1}}
    {-i\Omega\left(\Omega+\xi^i\sigma\right) + 
    \left(\gamma_{1,1}-\xi^r\sigma\right)}
    \cdot
    \frac{
    2\overline{a}^* \left(\eta\Gamma\right)^{3/2}
    }{
    i\left(\overline{\Delta}-\Omega\right) + \Gamma/2
    }\cdot
    s_p
\end{equation*}
where $\sigma={2G\beta\abs{\overline{a}}^2\overline{\Delta}}/
    ({\overline{\Delta}^2+\Gamma^2/4})$. From the above expression, the phase can be approximated as:
\begin{equation*}
    \phi(\Omega)\approx\arctan{
    \frac{
    \Omega\left(\gamma_{1,1}-\xi^r\sigma\right) -
    \gamma_{1,1}\left(\Omega+\xi^i\sigma\right)}
    {\Omega\left(\Omega+\xi^i\sigma\right) +
    \gamma_{1,1}\left(\gamma_{1,1}-\xi^r\sigma\right)}
    }
\end{equation*}
where we neglected the bare resonator phase response.\\ 
We will now assume the peak group delay to have a functional form analogous to Eq.~\eqref{eq:delay_beat}:
\begin{equation}
    \tau_g(\Omega\rightarrow0) \approx
    -\frac{\mathcal{V}}{\gamma_{eff}}
\end{equation}
\\
where $\gamma_{eff}$ is an effective decay rate that depends on the parameters $\gamma_{i,j}$.
With this hypothesis, and in the presence of linear dispersion at $\Omega\rightarrow0$, the asymptotic group delay can then be expressed as:
\begin{align*}
    \tau_g &=
    \lim_{\Omega\rightarrow0} -\frac{\phi}{\Omega}
    =
    -\frac{1}{\gamma_{1,1}}
    \left[
    1-\frac{\gamma_{1,1}}{\Gamma_{IT}}
    \left(1+\xi^i/\Omega\sigma\right)
    \right]\\
    &\stackrel{\sigma\rightarrow\infty}{\approx}
    -\frac{1}{\xi^r}\frac{d\xi^i}{d\Omega}=
    \tau_g^{min}
\end{align*}
\\
where in the last step we assumed $\Gamma_{IT}(0)\gg\gamma_{1,1}$, and thus $\mathcal{V}\rightarrow 1$ in the blue-detuning regime.
The final analytic result for the case $n=3$ is reported in the main text.


%

\end{document}